\documentclass[twocolumn]{aastex631}
\usepackage{graphicx} 
\usepackage{indentfirst}
\usepackage{amsmath}
\usepackage{mdsymbol}
\usepackage{gensymb}
\usepackage{xcolor}

\begin{document}

\title{Quasi-thermal Photosphere Emission from Structured Jets of Gamma-Ray Bursts} 

\author[0009-0002-7851-3706]{Ding-Fang Hu}
\affiliation{Purple Mountain Observatory, Chinese Academy of Sciences, Nanjing 210023, China}
\affiliation{School of Astronomy and Space Sciences, University of Science and Technology of China, Hefei 230026, China}

\author[0000-0001-9648-7295]{Jin-Jun Geng}\thanks{E-mail: jjgeng@pmo.ac.cn}
\affiliation{Purple Mountain Observatory, Chinese Academy of Sciences, Nanjing 210023, China}

\author[0000-0001-7892-9790]{Hao-Xuan Gao}
\affiliation{Purple Mountain Observatory, Chinese Academy of Sciences, Nanjing 210023, China}

\author[0000-0003-2553-2217]{Jing-Zhi Yan}\thanks{E-mail: jzyan@pmo.ac.cn}
\affiliation{Purple Mountain Observatory, Chinese Academy of Sciences, Nanjing 210023, China}

\author[0000-0002-6299-1263]{Xue-Feng Wu}\thanks{E-mail: xfwu@pmo.ac.cn}
\affiliation{Purple Mountain Observatory, Chinese Academy of Sciences, Nanjing 210023, China}
\affiliation{School of Astronomy and Space Sciences, University of Science and Technology of China, Hefei 230026, China}

\begin{abstract}
The prompt emission of gamma-ray bursts (GRBs) is supposed to be released from the relativistic jet launched from the central engine.
Apart from the non-thermal nature of the spectra in a majority of GRBs,
there is evidence for the presence of quasi-thermal components in the prompt emission of a few GRBs according to observations by \textit{Fermi} satellite.
On the other hand, the GRB jet has been revealed as structured in recent research.
The theoretical observed spectra of photosphere emissions by an off-axis observer and the dependence of the spectra on the viewing angle under the assumption of structured jets remain unexplored.
In this paper, we numerically calculate the instantaneous photosphere spectra by different viewing angles from a structured jet, from which relevant temporal and spectral characteristics are derived.
Moreover, we address the necessity of proper treatment of the outflow boundary in the photosphere emission scenario.
Furthermore, our calculations suggest that the \textit{Einstein Probe} and \textit{Space-based multi-band astronomical Variable Object Monitor} will have the capability to detect the short GRBs similar to GRB 170817A up to a luminosity distance of 200~Mpc if the off-axis viewing angle is less than $10\degree$.

\end{abstract}


\section{INTRODUCTION} \label{sec:intro}
Gamma-ray bursts (GRBs) are catastrophic events that are the most powerful explosions in the universe. Since its discovery \citep{Klebesadel_1973}, the radiation mechanism of prompt emission has remained a subject of debate. 
The spectra of GRBs could be phenomenally described by a smoothly joint broken power law, called ``Band Function'' \citep{Band_1993}, with a typical value of the spectral index in the low-energy regime being $\alpha \sim -1$ \citep{Preece_2000, Ghirlanda_2002, Nava_2011}. 
The broad non-thermal feature of the prompt emission favors the synchrotron origin from electrons accelerated by the dissipation processes like internal shock or magnetic reconnection~\citep{Rees_1994, Thompson_1994, Spruit_2001}.
The simple synchrotron emission scenario experienced challenges in explaining the characteristic value of $\alpha \sim -1$~\citep{Preece_1998}, although this concern has alleviated through detailed treatment of electron cooling processes ~\citep{Derishev_2001, Daigne_2011, Uhm_2014, ZhaoXH_2014, Geng_2018_apjs}.
Recently, it has been suggested that synchrotron spectra from electrons in evolving (fast-to-slow) cooling regimes can match $\sim95\%$ of all time-resolved spectra of the brightest long GRBs observed by the Fermi Gamma-Ray Observatory's gamma-ray burst monitor (GBM: 8~keV–40~MeV) by comparing the 
physical model with observed data directly \citep{Burgess_2020}.
Nevertheless, there is still presence of a clear thermal or quasi-thermal component in prompt emission of a few GRBs \citep{Abdo_2009, Ryde_2009, Ryde_2010, Axelsson_2012, moretti2013fermi, Burgess_2011, Guiriec_2011}, driving relevant explanations including phototsphere models \citep{Pe'er_2007, Pe'er_2011, Deng_2014, Meng_2018, Meng_2019, Meng_2022, Meng_2023}, Comptonized photosphere models \citep{Meszaros_2000, Rees_2005, Pe'er_2006, Beloborodov_2010, Ryde_2010, Vurm_2011}, and the radiation-mediated shock model \citep{Samuelsson_2023, Alamaa_2024}.

The approximation of a top-hat jet, i.e., a uniform jet with a sharp edge, was widely adopted in modeling both the prompt and afterglow emissions of GRBs in the past~\citep{Meszaros_1999, Panaitescu_1999}. Lately, extensive observations of GRB 170817A~\citep{Alexander_2017, Arcavi_2017,Mooley_2018,Lyman_2018,Kasliwal_2017} suggest that the jet should be structured, which is also supported by theories~\citep{Lamb_2017,Troja_2017,Troja_2019,Meng_2018,Granot_2017,Granot_2018,Lazzati_2018,Kathirgamaraju_2018,Geng_2019,Li_2019,Ryan_2020}, motivating theorists to revisit both prompt and afterglow models that involve relativistic jets.
In general, a structured jet consists of two parts, an inner narrow jet core and an outer continuous jet wing or a mildly relativistic cocoon component in theory~\citep{Dai_2001,Zhang_2002,Kumar_2003,Rossi_2002,Perna_2003}.
Quasi-thermal emissions are expected a short time after the cocoon shock breakout when matter and radiation come into equilibrium \citep{Gottlieb_2018}.   
For specific, both synchrotron and photosphere emission could interpret the spectral feature of GRB 170817A at a comparable fitting 
goodness~\citep{Meng_2018}. 
The quasi-thermal photosphere emission from a realistic structured jet hence remained to be further explored, especially for short GRBs with larger jet opening angles in the wide field survey era.

This article will first introduce the basic assumptions of the photosphere emission and the geometric relationship of off-axis observations in Section \ref{sec:bas_mod}. In Section \ref{sec:mod&res}, we calculate the instantaneous spectra and explore their dependence on various viewing angles and different treatments of boundary conditions. We also calculate the theoretical instantaneous spectrum of short GRBs similar to GRB 170817A and analyze the feasibility of photosphere emission off-axis observations in combination with the sensitivity of the \textit{Einstein Probe Wide-field X-ray Telescope} (EP-WXT), \textit{ECLAIRs on board Space-based multi-band astronomical Variable Object Monitor} (SVOM-ECLAIRs) and \textit{Burst Alert Telescope on Neil Gehrels Swift Observatory} (Swift-BAT)~\citep{Yuan_2022, Zhang_2022, Paul_2011}. Finally, in Section \ref{sec:conclu}, we present some discussions, and the conclusions are drawn.

\section{PHOTOSPHERE EMISSIONS OF A STRUCTURED JET}\label{sec:bas_mod} 
\subsection{Structured Jets}
\label{subsec:stru_jets}
\begin{figure}
    \centering
    \includegraphics[width=0.5\textwidth]{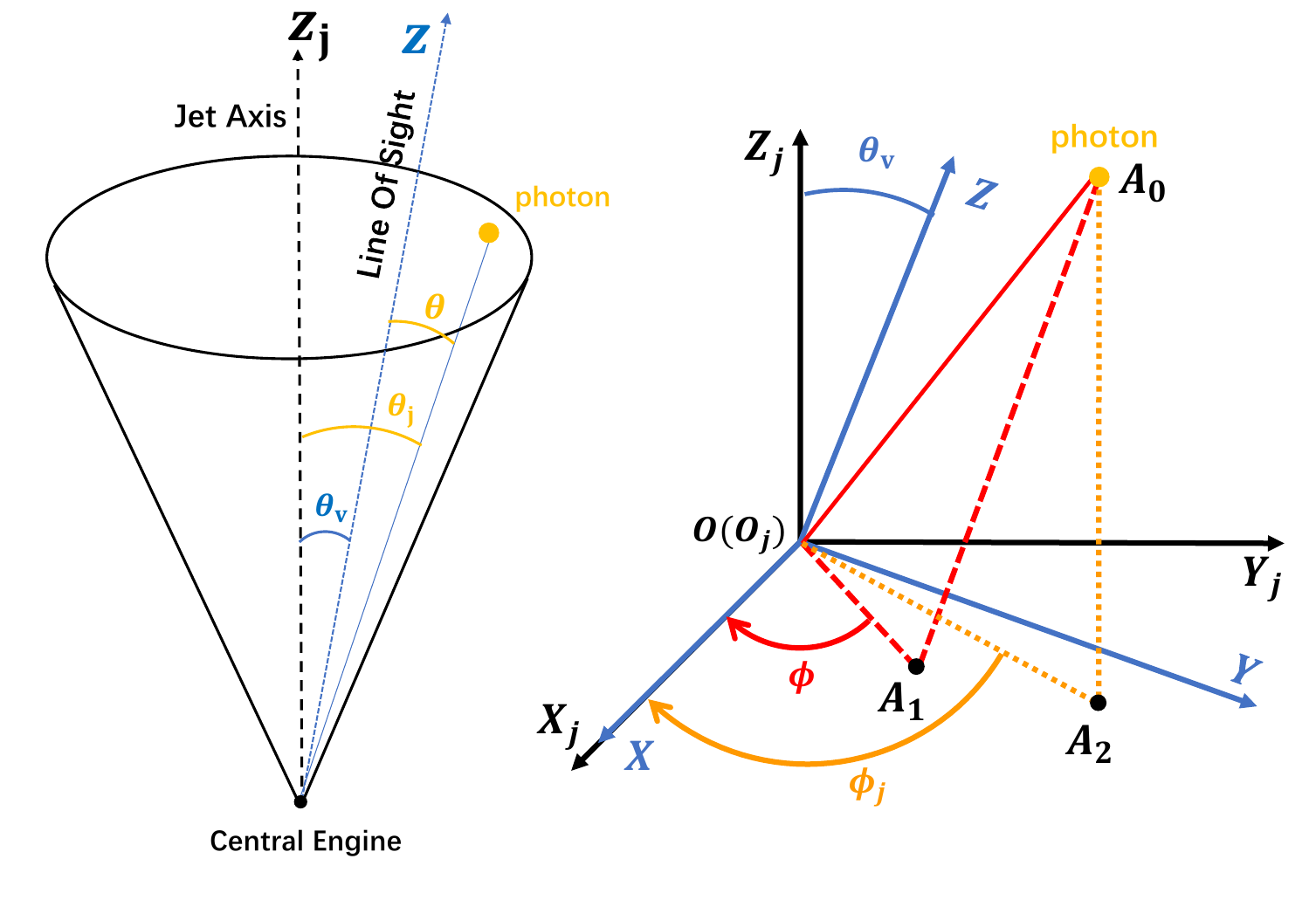}
    \caption{Schematic diagram of the structured jet. $\theta_{\mathrm{v}}$ represents the viewing angle, which is also the angle between the observer frame and the jet frame. $A_{1}$ and $A_{2}$ represents the projection of photon $A_{0}$ into the XOY plane and $X_{\mathrm j}Y_{\mathrm j}Z_{\mathrm j}$ plane. The polar angles and the azimuthal angles are denoted as $\theta$ ($\theta_{\mathrm{j}})$ and $\phi$ ($\phi_{\mathrm{j}}$), respectively. }
    \label{fig:stru_jet}
\end{figure}

During the prompt phase, the central engine injects a continuous stream of photons and plasma into the space. When newly injected photons encounter electrons, they undergo Thomson scattering. The photons diffuse outward along the radial direction until the last scattering and then escape from the photosphere in a straight line towards the observer \citep{Pe'er_2011}. To facilitate calculations, we will use two coordinate systems: the jet coordinate system and the observer coordinate system. Figure \ref{fig:stru_jet} shows the geometric relationship between the two coordinate systems. The polar angle and the azimuthal angle are denoted as $\theta_{\mathrm j}$, $\phi_{\mathrm j}$ in the jet system, $\theta$, $\phi$ are measured in the observer system, and the viewing angle is $\theta_{\mathrm v}$ \citep{Fan_2008, Geng_2018_apj}. The geometry relation gives that
\begin{equation}
    \theta_{\mathrm j}=\arccos[\cos(\theta)\cos(\theta_{\mathrm v})+\sin(\theta)\sin(\theta_{\mathrm v})\cos(\phi)].
    \label{eq:co_trans}
\end{equation}
For the structured jet, the bulk Lorentz factor distribution of the plasma and the equivalent isotropic luminosity distribution of photons could be assumed to be power law, i.e.,
\begin{align}
    L(\theta_{\mathrm{j}})=
    \begin{cases}
    L_{\mathrm c}  & \text{if} \ \theta_{\mathrm{j}} \leq \theta_{\mathrm c}, \\
    L_{\mathrm c}(\theta_{\mathrm{j}} /\theta_{\mathrm c})^{-k_{\mathrm e}} & \text{if} \ \theta_{\mathrm c}<\theta_{\mathrm{j}} \leq \theta_{\mathrm m},
    \end{cases} \\
        \Gamma(\theta_{\mathrm{j}})=
    \begin{cases}
    \Gamma_{\mathrm c}  & \text{if} \ \theta_{\mathrm{j}} \leq \theta_{\mathrm c}, \\
    \Gamma_{\mathrm c}(\theta_{\mathrm{j}} /\theta_{\mathrm c})^{-k_{\mathrm\Gamma}}+1 & \text{if} \ \theta_{\mathrm c}<\theta_{\mathrm{j}} \leq \theta_{\mathrm m},
    \end{cases}
\end{align}
where $\theta_{\mathrm c}$ is the half-opening angle of the inner core, $\theta_{\mathrm m}$ is the edge of the outer jet, and $L_{\mathrm c}$, $\Gamma_{\mathrm c}$ are the luminosity and Lorentz factor of the inner core respectively.

The temperature at point $(r,\theta)$ in the observer frame is related to the temperature distribution of the photosphere by
\begin{equation}
    T(r,\theta)=DT^{\prime} (r,\theta_{\mathrm j}),
\end{equation}
where the Doppler factor $D$ is expressed by
\begin{equation}
    D=[\Gamma(\theta_{\mathrm j})(1-\beta(\theta_{\mathrm j})\cos(\theta))]^{-1}.
\end{equation}
Hereafter, the superscript prime ($\prime$) is used to denote the quantities in the comoving frame. 
Thus, the comoving temperature distribution of the photosphere is written as \citep{Meszaros_2000, Rees_2005, Pe'er_2010}
\begin{align}
    &T^{\prime}(r,\theta_{\mathrm j})=T(r,\theta=0)/2\Gamma(\theta_{\mathrm j})= \notag \\
    &\left\{
    \begin{array}{l}
      T_{0}(\theta_{\mathrm j})/(2\Gamma (\theta_{\mathrm j})), \text{\ \ \ \ \ \ \ \ \ \ \ \ \ \ \ \ \ } r<r_{\mathrm s}(\theta_{\mathrm j})<r_{\mathrm{ph}}(\theta_{\mathrm j}), \\
      T_{0}(\theta_{\mathrm j})(r/r_{\mathrm s})^{-2/3}/(2\Gamma(\theta_{\mathrm j})), \text{\ \ \ \ \ } r_{\mathrm s}(\theta_{\mathrm j})<r<r_{\mathrm{ph}}(\theta_{\mathrm j}),\\
      T_{0}(\theta_{\mathrm j})(r_{\mathrm{ph}}/r_{\mathrm s})^{-2/3}/(2\Gamma(\theta_{\mathrm j})), \text{\ \ } r_{\mathrm s}(\theta_{\mathrm j})<r_{\mathrm{ph}}(\theta_{\mathrm j})<r, \label{eq:tem}      
    \end{array}
    \right.
\end{align}
and
\begin{align}
     T_{0}(\theta_{\mathrm j})=(\frac{L(\theta_{\mathrm j})}{4\pi r_{0}^{2}ca})^{\frac{1}{4}},\label{eq:cen-tem}\\ 
    r_{\mathrm s}(\theta_{\mathrm j})=\eta(\theta_{\mathrm j}) r_{0}=\Gamma(\theta_{\mathrm j}) r_{0},\label{eq:rs}\\
    \int_{r_{\mathrm{ph}}}^{\rm observer} d\tau=1,\label{eq:rph}   
\end{align}
where $T_{0}$ is the temperature at the basis $r_{0}$ of the outflow, $r_{\mathrm s}$ and $r_{\mathrm{ph}}$ are the saturation radius and photosphere radius \citep{Daigne_2002}, and $\eta$ is the dimensionless entropy of the photosphere. 
It is worth noting that the discussed temperature distribution here
holds for the regime of $r_{\mathrm s}<r_{\mathrm{ph}}$ \citep{Rees_1994, Thompson_1994, Pe'er_2007}, so we have $\eta(\theta_{\rm j})=\Gamma(\theta_{\rm j})$.
Additionally, due to the anisotropy of photon energy distribution, the central temperature $T_{0}$ is also asymmetrical and depends on $\theta_{\mathrm j}$.

\subsection{Calculation of Spectra}
\label{subsec:cal_spec}
\begin{figure}
    \centering
    \includegraphics[width=0.5\textwidth]{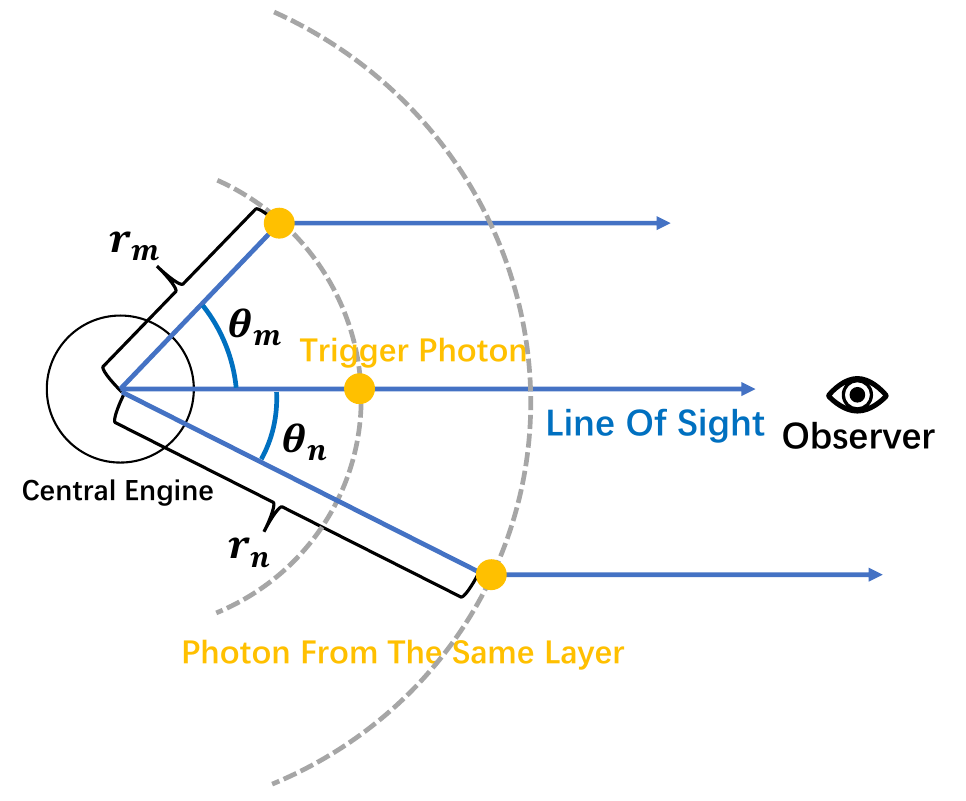}
    \caption{A geometric illustration of time delay. Photons injected at the same moment escape at different radii and latitudes, resulting in them arriving at the observer at different times.}
    \label{fig:t_geo}
\end{figure}
We use the term $\hat{t}$ to represent the central engine time and set the time zero point ($\hat{t}=0$) as the moment of the GRB explosion, and use $t_{\mathrm{obs}}$ to represent the observer time, with the zero point defined as the moment the observer receives the first photon.

We consider photons injected within the time interval $\hat{t}-\hat{t}+d\hat{t}$ as a layer. For photons from the same layer, they escape the photosphere from different latitudes and radii, causing them to reach the observer at different $t_{\mathrm{obs}}$ \citep{Pe'er_2008}. We call the photon that reaches the observer first in each layer the ``trigger photon". Combining Figure \ref{fig:t_geo}, the time difference between the trigger photon and other photons arriving at the observer can be expressed by $\Delta t=ru/\beta(\theta_{\mathrm j}) c$, where $u=1-\beta(\theta_{\mathrm j}) \cos(\theta)$. Thus, for photons arriving at the observer at $t_{\mathrm{obs}}$, the injection time $\hat{t}$ and the time delay $\Delta t$ due to geometric effects should satisfy:
\begin{equation}
    \frac{t_{\mathrm{obs}}}{1+z}=\hat{t}+\frac{ru}{\beta(\theta_{\mathrm j}) c}. \label{eq:t-rel}
\end{equation}
The contribution of the photons injected at engine time $\hat{t}$ to the instantaneous specific flux observed by the observer at $t_{\mathrm{obs}}$ and $\theta_{\mathrm{v}}$ can be expressed as \citep{Zhang_2018,Deng_2014}
\begin{align}
    &\hat{F}_{\nu_{\mathrm{obs}}}(\nu_{\mathrm{obs}},t_{\mathrm{obs}},\theta_{\mathrm{v}},\hat{t})=&& \notag\\
    &\frac{1+z}{{d_{\mathrm L}^{2}}} \iint \frac{d\dot{N}}{d \Omega}P(r,\Omega)R(\nu,T)h\nu
    \delta(t-(\hat{t}+\frac{ru}{\beta c})) d\Omega dr, && \label{eq:lay-contr}
\end{align}
where $\nu=(1+z)\nu_{\mathrm{obs}}$, $t=t_{\mathrm{obs}}/(1+z)$, $d\dot{N}/d\Omega$ denotes the number of photons injected per unit time per unit solid angle by the central engine and can be calculated by
\begin{equation}
    \frac{d\dot{N}(\theta_{\mathrm j})}{d\Omega}=\frac{L(\theta_{\mathrm j})}{4 \pi k T_{0}(\theta_{\mathrm j})}, \label{eq:N-distr}
\end{equation}
$R(\nu,T)$ is the ratio of a photon with frequency $\nu$ in a Plank distribution with a temperature of $T$, 
\begin{equation}
    R(\nu,T)=\frac{h^{3}\nu^{2}}{2.404(kT)^{3}[e^{\frac{h\nu}{kT}}-1]},
\end{equation}
and $P(r,\Omega)$ denotes the probability density that the last scattering occurs at $(r, \Omega)$. According to the probability density function to describe the last scattering event of spherically symmetric jets  \citep{Pe'er_2008, Beloborodov_2011, Deng_2014}, the probability function for the structured jet written as
\begin{equation}
    P(r,\Omega)=P(r,\theta,\phi)=\frac{D^{2}(\theta,\phi)\sigma_{\mathrm T}n(r,\theta,\phi)e^{-\tau}}{8\pi\Gamma^{2}(\theta,\phi) A}.\label{eq:prob}
\end{equation}
The proportionality factor $A$ will be discussed in Section \ref{subsec:con&inf}. It is important to note that due to the jets are not symmetrical to an off-axis observer, this probability density also depends on the azimuthal angle $\phi$,
which is further derived in Appendix \ref{app:prob}.

\section{MODELS AND NUMERICAL RESULTS} 
\label{sec:mod&res}
Using the formulations above, we then calculate the spectral temporal evolution 
of photosphere emissions of structured jets and explore its dependence on the viewing angles. Three representative models are considered successively.

\subsection{Constant Luminosity With Infinite Outer Boundary}
\label{subsec:con&inf}

As a simple case, we assume that the luminosity of the central engine $L_{\mathrm{c}}(\hat{t})$ is a constant and consider that the outer boundary of the photosphere is infinite. In the probability density function of the last scattering event, the term $\tau$ represents the optical depth experienced by the photon from the location of the last scattering to the observer. 
The infinite outer boundary (along the $r$ direction) assumption of the jet means that the optical depth experienced by the photon after scattering at $(r_{1},\theta_{1},\phi_{1})$ before escaping is \citep{Abramowicz_1991, Pe'er_2008, Beloborodov_2011}
\begin{equation}   \tau=\tau(r_{1},\theta_{1},\phi_{1})=\int_{r_{1}}^{\infty}D^{-1}\sigma_{\mathrm T}n^{\prime}\frac{dr}{\cos(\theta)},
\label{eq:tau_con&inf}
\end{equation}
where 
\begin{equation}
    n^{\prime}(\theta_{\mathrm j})
    =\frac{\dot{M}(\theta_{\mathrm j})}{4\pi m_{\mathrm p}\beta(\theta_{\mathrm j})c\Gamma(\theta_{\mathrm j})r^{2}}, 
\end{equation}
\begin{equation}
    \dot{M}(\theta_{\mathrm j})=\frac{L(\theta_{\mathrm j})}{\eta(\theta_{\mathrm j})c^{2}},   
\end{equation}
and the relationship between $\theta$ and $\theta_{\mathrm j}$ is given in Equation \ref{eq:co_trans}. Equation \ref{eq:tau_con&inf} does not have an analytical solution and requires numerical calculation methods.

Furthermore, in the calculation of factor $A$, we no longer rely on probability normalization but instead consider photon number conservation. That is, the number of photons injected from the central engine is equal to the number of photons escaping from the photosphere
\begin{align}
    &\int\int \cdot [\int\int\frac{d\dot{N}(\theta,\phi)}{d \Omega} P(r,\Omega)drd\Omega]\cdot d\Omega_{\mathrm v}d\hat{t}&&\notag\\  
    &=\int\int \frac{d\dot{N}(\theta,\phi)}{d \Omega}\cdot d\Omega d\hat{t},&&
\end{align}
where $\frac{d\dot{N}(\theta,\phi)}{d \Omega}$ is determined in Equation \ref{eq:N-distr} in the observer frame, and $d\Omega_{\mathrm v}$ represents the solid angle subtended by the engine to the observer. Hence the proportionality constant $A$ can be calculated by
\begin{equation}
    A=\frac{\int\int\cdot[\int\int\frac{d\dot{N}(\theta,\phi)}{d \Omega}\cdot\frac{\sigma_{\mathrm T}nD^{2}e^{-\tau}}{8\pi\Gamma^{2}}drd\Omega]\cdot d\Omega_{\mathrm v} d\hat{t}}{\int\int\frac{d\dot{N}(\theta,\phi)}{d \Omega}\cdot d\Omega d\hat{t} }.\label{eq:ph_cons}
\end{equation}

The photons received by the observer at a certain time $t_{\mathrm{obs}}$ come from different layers, with each layer corresponding to an engine time $\hat{t}$.
The total observed instantaneous specific flux at $t_{\rm obs}$ can be obtained by integrating the contribution in each layer in Equation \ref{eq:lay-contr}, i.e., 
\begin{equation}
    {F}_{\nu_{\mathrm{obs}}}(\nu_{\mathrm{obs}},t_{\mathrm{obs}},\theta_{\mathrm{v}})=\int_{0}^{\frac{t_{\mathrm{obs}}}{1+z}} \hat{F}_{\nu_{\mathrm{obs}}}(\nu_{\mathrm{obs}},t_{\mathrm{obs}},\theta_{\mathrm{v}},\hat{t})d\hat{t},
    \label{eq:lay_int}
\end{equation}
where ${F}_{\nu}$ represents the observed specific flux and $\hat{F}_{\nu}$ represents the contribution of the specific flux per time
for the photons injected at engine time $\hat{t}$ (see Equation \ref{eq:lay-contr}).

\begin{figure*}[!htb]
    \centering
    \includegraphics[width=\textwidth]{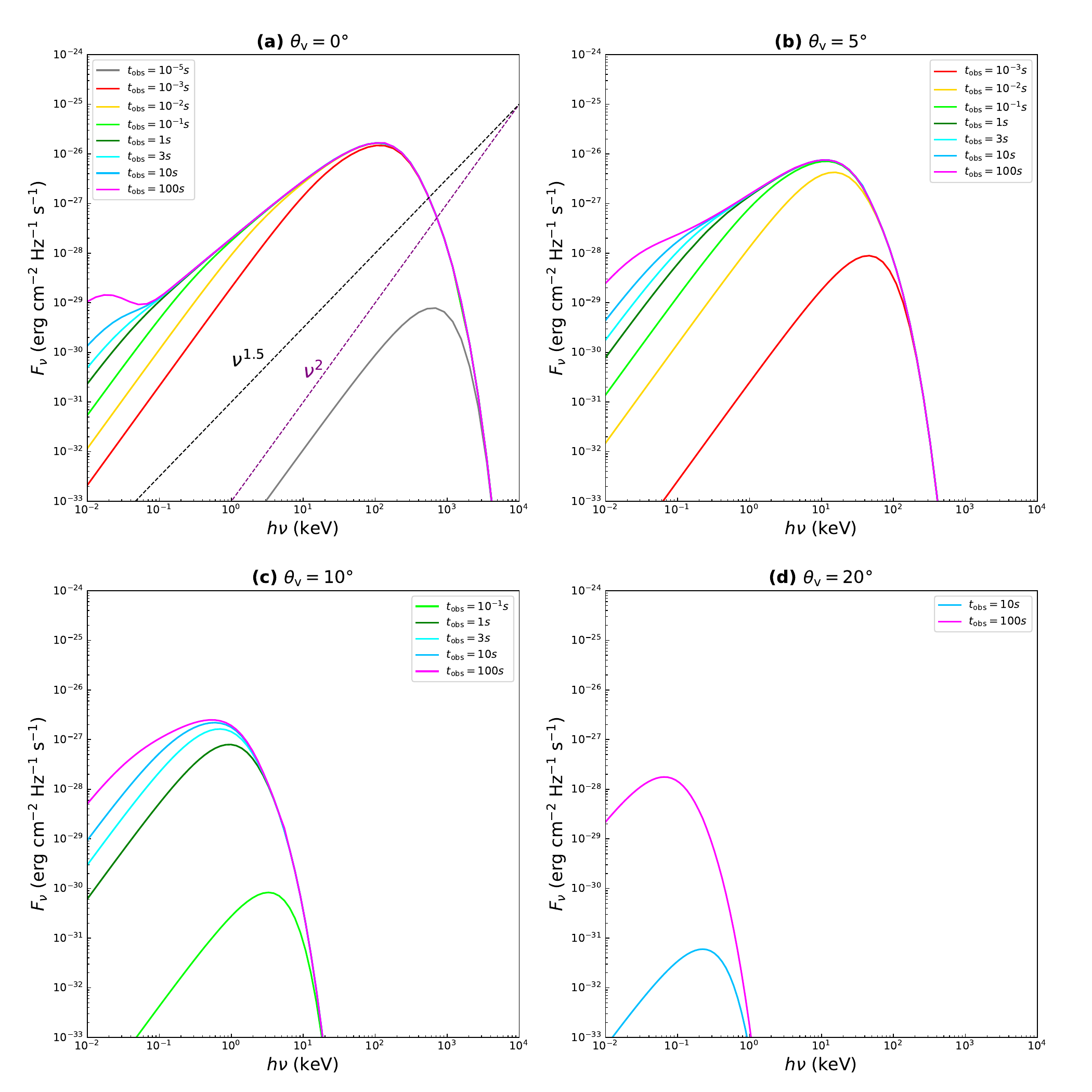}
    \caption{Instantaneous photosphere spectra with constant luminosity and infinite outer boundary by different viewing angles $\theta_{\mathrm v}$, the inner-core Lorentz factor $\Gamma_{\mathrm c}=300$, luminosity $L_{\mathrm c}=10^{52}~\mathrm{erg\:s^{-1}}$, distance $d_{\mathrm L}=8.8\times10^{27}~\mathrm{cm}$ ($z\sim 0.5$), half-opening angle of the inner-core $\theta_{\mathrm c}=3\degree$, the maximum half-opening angle of the jet $\theta_{\mathrm m}=30\degree$, index values are $k_{\mathrm{\Gamma}}=2$, $k_{\mathrm e}=2$. Different colors stand for different $t_{\mathrm{obs}}$, $10^{-5}~\mathrm s$ (grey), $10^{-3}~\mathrm s$ (red), $10^{-2}~\mathrm s$ (gold), $10^{-1}~\mathrm s$ (lime), $1~\mathrm s$ (green), $3~\mathrm s$ (cyan), $10~\mathrm s$ (deepskyblue), $100~\mathrm s$ (magenta), and the dashed lines are the reference line for $F_{\nu}\propto\nu^{2}$ (purple), $F_{\nu}\propto\nu^{1.5}$ (black). For off-axis observations, the magnitude of instantaneous photosphere spectra for early time is very low, so part of the early spectra in Figure (b) (c) (d) are not drawn.}
    \label{fig:spec_con&inf}
\end{figure*}
In the numerical results as shown in Figure \ref{fig:spec_con&inf}, we found that the shape of the early instantaneous spectrum at the low-frequency range below the peak energy is close to the blackbody spectrum under the Rayleigh-Jeans limit, which is proportional to $\nu^{2}$. Due to the different temperatures at different latitudes in the photosphere and the fact that photons from higher latitudes will reach the observer at a later time, the spectral index will decrease slightly and approaches 1.5 gradually ($F_{\nu}\propto \nu^{1.5}$),
which is a result of the superposition of blackbody spectra with multi-temperatures. 
Since photons from higher latitudes with lower energy will be observed by a larger time delay, a bump structure appears in the low-frequency range at $\sim 100$~s as shown in Figure~\ref{fig:spec_con&inf} (a).
On the other hand, the peak energy of the spectrum decreases monotonically with time since the early radiation originates from regions with smaller radii corresponding to higher temperatures (Equation \ref{eq:tem}).

Comparing the results under different viewing angles, it's obvious that the flux density and peak energy from the on-axis direction are higher than that from the off-axis direction. This is because the on-axis direction has higher luminosity with smaller optical depth and higher photosphere temperature.
In addition, the evolution rate of the instantaneous spectrum is $\theta_{\rm v}$ dependent. For the on-axis case, the spectrum evolves to a quasi-saturation shape very soon ($\sim 10^{-3}$~s). As $\theta_{\rm v}$ increases, the evolving timescale becomes larger, e.g., the spectrum changes significantly from 10 to $10^2$~s for $\theta_{\rm v} = 20\degree$. The evolution rate is mainly caused by the angular dependence of the photosphere radius $r_{\mathrm{ph}}$. Combined with the probability density function (Equation \ref{eq:prob}), most photons will undergo their last scattering near the photosphere radius, these photons arrive at a delayed time after the ``trigger photon'', which is given by
\begin{equation}
    \Delta t_{\mathrm{D}}(\theta_{\mathrm{v}})\gtrsim(1+z)\frac{r_{\mathrm{ph}}(\theta_{\mathrm{v}})}{2c\Gamma^2(\theta_{\mathrm{v}})},
\end{equation}
where the equality corresponds to photons injected right along the line of sight (LOS), and the photons from higher latitudes (with respect to the LOS) are observed at a larger time delay. On the other hand, the photosphere radius could be approximated as
\begin{align}
    &r_{\mathrm{ph}}(\theta_{\mathrm{v}})\simeq  \frac{L(\theta_{\mathrm{v}})\sigma_{\mathrm{T}}}{8\pi m_{\mathrm{p}}c^3\Gamma^3(\theta_{\mathrm{v}})}\simeq\notag\\
    &\left\{
    \begin{array}{l}
    5.9\times10^{12}(\frac{L_{\mathrm{c}}}{10^{52}})(\frac{\Gamma_{\mathrm{c}}}{100})^{-3}~\mathrm{cm}, \text{\ \ } \theta_{\mathrm{v}}\leq\theta_{\mathrm{c}}, \\
    5.9\times10^{12}(\frac{L_{\mathrm{c}}}{10^{52}})(\frac{\Gamma_{\mathrm{c}}}{100})^{-3}(\frac{\theta_{\mathrm{v}}}{\theta_{\mathrm{c}}})^{-k_{\mathrm{e}}+3k_{\Gamma}}~\mathrm{cm}, \text{\ \ } \theta_{\mathrm{c}}<\theta_{\mathrm{v}}\leq\theta_{\mathrm{m}}.
    \end{array}
    \right.\label{eq:rph_appro}
\end{align}
So the time delay is written as
\begin{align}
    &\Delta t_{\mathrm{D}}(\theta_{\mathrm{v}})\gtrsim\notag\\
    &\left\{
    \begin{array}{l}
    9.8\times10^{-3}(1+z)(\frac{L_{\mathrm{c}}}{10^{52}})(\frac{\Gamma_{\mathrm{c}}}{100})^{-5}~\mathrm{s}, \text{\ \ } \theta_{\mathrm{v}}\leq\theta_{\mathrm{c}},\\
    9.8\times10^{-3}(1+z)(\frac{L_{\mathrm{c}}}{10^{52}})(\frac{\Gamma_{\mathrm{c}}}{100})^{-5}(\frac{\theta_{\mathrm{v}}}{\theta_{\mathrm{c}}})^{-k_{\mathrm{e}}+5k_{\Gamma}}~\mathrm{s}, \text{\ \ } \theta_{\mathrm{c}}<\theta_{\mathrm{v}}\leq\theta_{\mathrm{m}}.   
    \end{array}
    \right.
\end{align}
The radiation radius grows and converges to the photosphere radius as the spectrum evolves to a quasi-saturation state, so the saturation timescale is comparable with $\Delta t_{\rm D}$. We could then expect the saturation timescale for four representative cases in Figure \ref{fig:spec_con&inf} to be $\sim 10^{-4}$~s, $\sim 10^{-3}$~s, $\sim 1$~s, and $\sim 10^2$~s respectively. 

\subsection{Constant Luminosity With Finite Outer Boundary} \label{subsec:con&fin}
In this model, we still assume that the luminosity of the central engine remains constant, but the outer boundary of the outflow is taken as finite. For an outflow suddenly released, its physical outer boundary at $\hat{t}$ can be expressed as
\begin{equation}
    r_{\mathrm{out}}(\theta_{\mathrm j},\hat{t})=\beta(\theta_{\mathrm j})c\hat{t}.
\end{equation}
With finite values for $r_{\mathrm{out}}$, photons escaping from the photosphere will not encounter any electrons after they catch up with the outer boundary.
The upper limit of integration in calculating the optical depth (Equation \ref{eq:catch-up}) is now dynamically determined rather than infinity.

\begin{figure}
     \centering
     \includegraphics[width=0.5\textwidth]{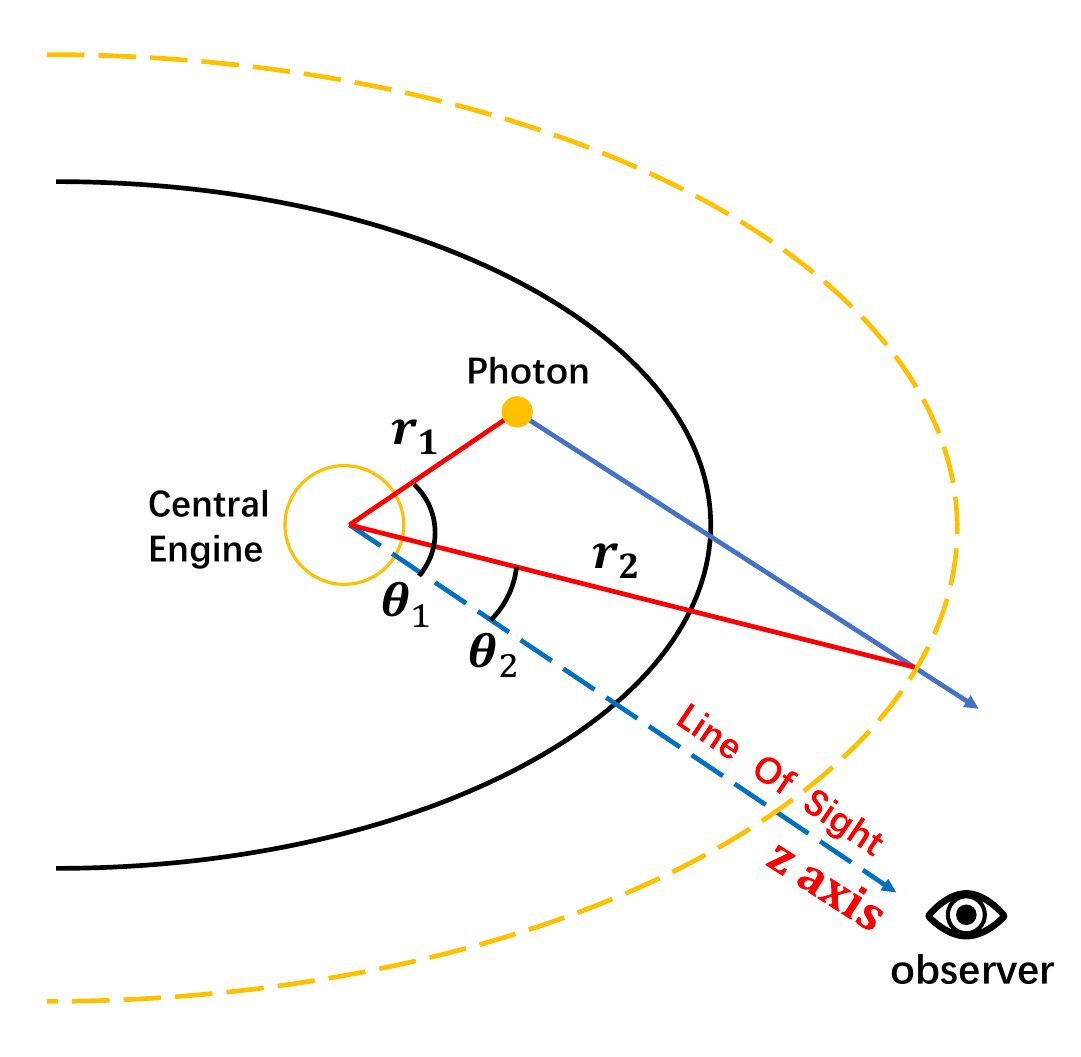}
     \caption{Geometric relationship of the catch-up process. The photon escapes from the photosphere at $(r_{1},\theta_{1})$, at this moment, the outer boundary is located at the solid black line, and then the photon catch-up the outer boundary at $(r_{2},\theta_{2})$ on the dashed yellow line.}
     \label{fig:cat_up}
 \end{figure}

Figure \ref{fig:cat_up} illustrates the geometric picture of the escaping photons catch-up the outer boundary of outflow, from which we have geometric relationships of
 \begin{equation}
\begin{cases}
r_{1}\sin(\theta_{1})=r_{2}\sin(\theta_{2}),\\
r_{2}\cos(\theta_{2})-r_{1}\cos(\theta_{1})=c\Delta t,\\
r_{2}=\beta(\theta_{\mathrm{j2}})c[\hat{t}+\frac{r_{1}}{\beta(\theta_{\mathrm{j1}})}c+\Delta t],
\label{eq:catch-up}
\end{cases}
\end{equation}
where $\theta_{\mathrm{j1}}$ and $\theta_{\mathrm{j2}}$ are the polar angles in the jet coordinate system, and $\Delta t$ represents the time required for the photon to escape from the photosphere and catch up with the outer boundary. Then we can solve out $r_{2}$ numerically, the upper limit of our integral for the optical depth $\tau$, i.e.,
\begin{equation}
    \tau=\int_{r_{1}}^{r_{2}}D^{-1}\sigma_{\mathrm T}n^{\prime}\frac{dr}{\cos(\theta)}.
    \label{eq:tau_con&fin}
\end{equation}
The specific algorithm for calculating optical depth can be found in the Appendix \ref{app:cal_tau} in detail.

Contrary to the probability density function that merely depends on $(r,\Omega)$ in Section \ref{subsec:con&inf}, photons injected at different $\hat{t}$ will catch up with the outer boundary at different locations here, leading to the dependence of the probability density function on the injection time $\hat{t}$,
\begin{equation}
    \hat{P}(r,\Omega,\hat{t})=\frac{D^{2}\sigma_{\mathrm T}\hat{n} e^{-\hat{\tau}}}{8\pi\Gamma^{2}A}.\label{eq:t-prob}
\end{equation}
Here we use ``\verb|^|'' to indicate that the function depends on injection time $\hat{t}$. Therefore, for different layers injected at different $\hat{t}$, we need to calculate the probability density function of them separately and compute the corresponding normalization constant $A$.

 \begin{figure*}
    \centering
    \includegraphics[width=\textwidth]{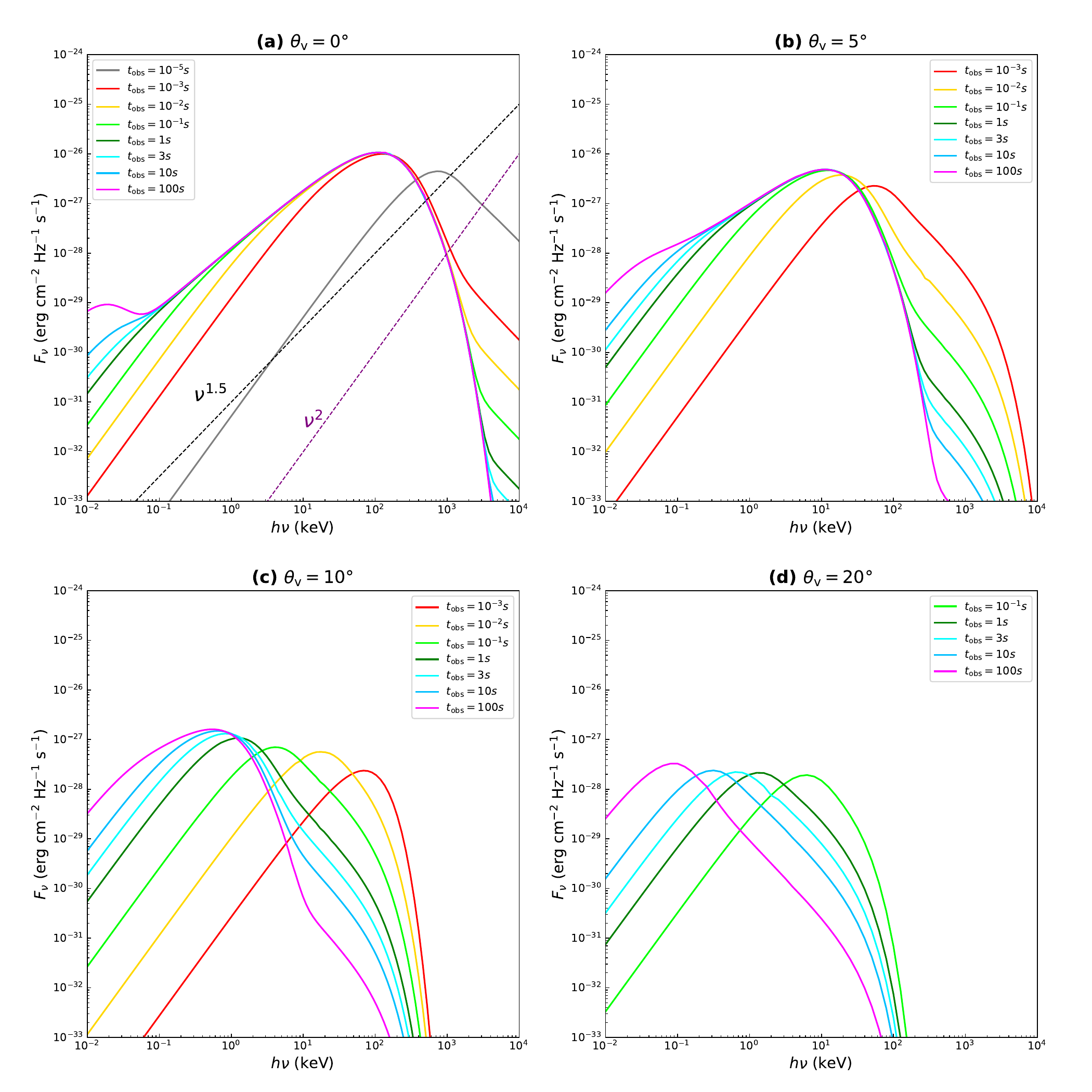}
    \caption{Instantaneous photosphere spectra with constant luminosity and finite outer boundary by different viewing angles $\theta_{\mathrm v}$. Parameters are the same as the Figure \ref{fig:spec_con&inf}, different colors stand for different $t_{\mathrm{obs}}$, $10^{-5}~\mathrm s$ (grey), $10^{-3}~\mathrm s$ (red), $10^{-2}~\mathrm s$ (gold), $10^{-1}~\mathrm s$ (lime), $1~\mathrm s$ (green), $3~\mathrm s$ (cyan), $10~\mathrm s$ (deepskyblue), $100~\mathrm s$ (magenta), and the dashed line is the reference line for $F_{\nu}\propto\nu^{2}$ (purple), $F_{\nu}\propto\nu^{1.5}$ (black).}
    \label{fig:spec_con&fin}
\end{figure*}

The instantaneous spectra of this model are shown in Figure \ref{fig:spec_con&fin}. 
We found that the shape of the instantaneous spectrum at the low-frequency range is still proportional to $\nu^{2}$ at early times, and the index gradually approaches 1.5 later, similar to those under the assumption of the infinite boundary in Section \ref{subsec:con&inf}. Meanwhile, the evolving rate depends on the viewing angle, the larger the viewing angle, the slower it evolves.
It is worth noting that the early flux density is several orders of magnitude higher than those in the infinite boundary model.
This is because the early radiation originates from a smaller radius within the photosphere. In the case of infinite boundary, photons from the early radiation would experience a non-physical large optical depth, resulting in a lower flux density. However, in the case of a finite boundary, photons of the early radiation catch up with the outer boundary quickly and do not experience a significant optical depth, resulting in a higher flux density.

In addition, the instantaneous spectra obtained with a finite boundary deviate from the pure exponential cut-off due to a ``hard'' component in the high frequencies, and it vanishes gradually with time. This power-law-like component is the result of thermal radiation with the evolving $r_{\mathrm{ph}}$. Compared with the constant $r_{\mathrm{ph}}$ (Equation \ref{eq:rph}) with an infinite boundary, $r_{\mathrm{ph}}$ now becomes progressively larger with the expansion of the outer boundary. As a smaller $r_{\mathrm{ph}}$ leads to a higher temperature (Equation \ref{eq:tem}) and the harder thermal radiation, it results in the hard component appearing in the high frequencies. Meanwhile, as the outer boundary of the photosphere expands, $r_{\mathrm{ph}}$ gradually increases and converges to the case of the infinite boundary, so this component gradually disappears.

Since the decreasing rate of the peak spectral energy is much slower for the off-axis case than that of the on-axis case, the hard component would last for a relatively longer time for the off-axis case. For example, in the case of the viewing angle of $20\degree$ in Figure \ref{fig:spec_con&fin}, 
the hard component still exists at 100 s, making the spectral curvature flatter than that of a simple plank function. It further illustrates the influence of different assumptions of the boundaries on the numerical results of instantaneous spectral evolution within the photosphere emission scenario, especially for the off-axis case.

\subsection{Variable Luminosity With finite Outer Boundary} \label{subsec:var&fin}
\begin{figure*}
    \centering
    \includegraphics[width=\textwidth]{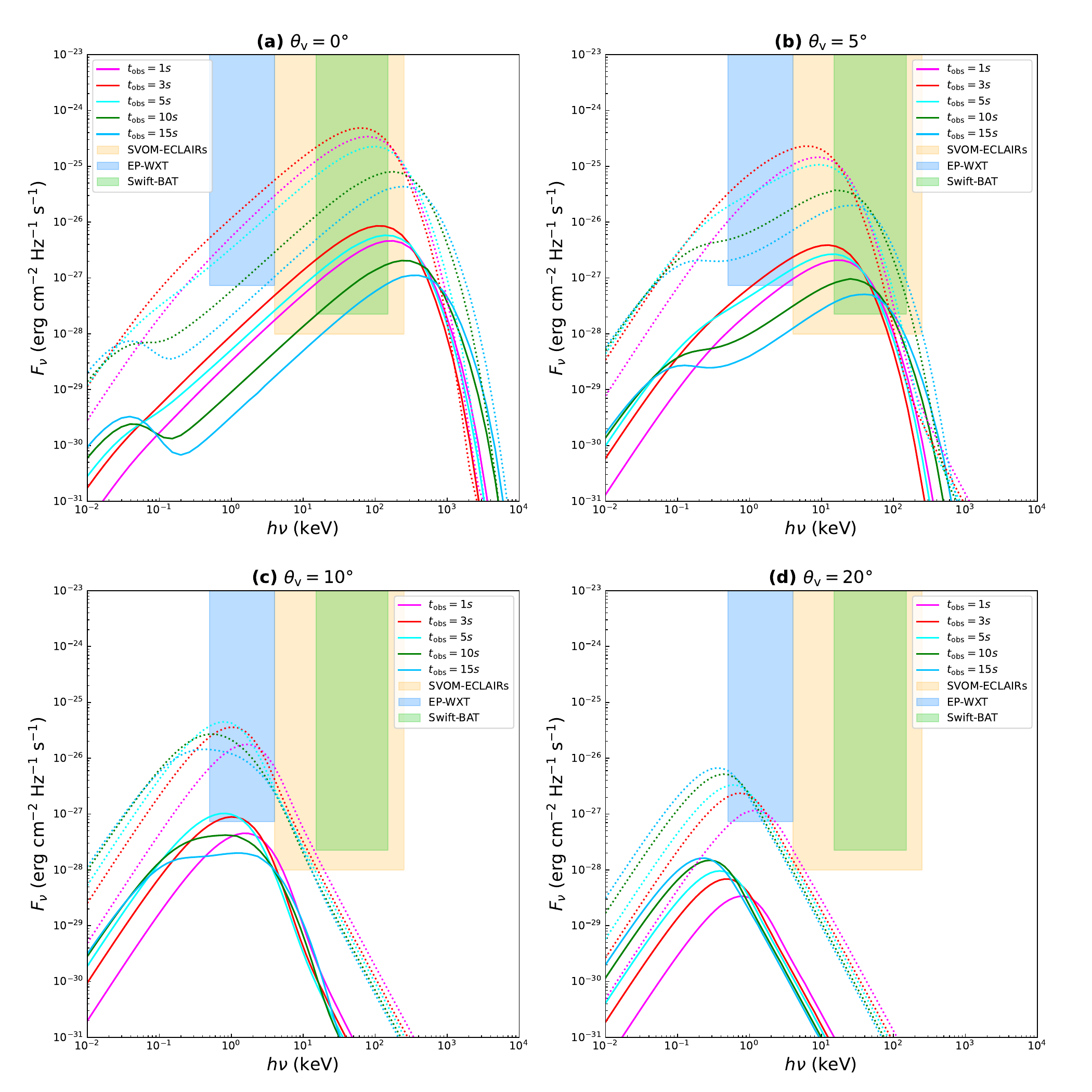}
    \caption{Instantaneous photosphere spectra with variable luminosity and finite outer boundary by different viewing angles $\theta_{\mathrm v}$. The solid line spectrum corresponds to a peak luminosity of $L_{\mathrm{cp}}=10^{52}~\mathrm{erg\:s^{-1}}$, and a luminosity distance of $d_{\mathrm{L}}=8.8\times10^{27}$~cm $(z\sim0.5)$. While the dashed line spectrum corresponds to a peak luminosity of $L_{\mathrm{cp}}=10^{53}~\mathrm{erg\:s^{-1}}$ and a luminosity distance of $d_{\mathrm{L}}=1.4\times10^{27}$~cm $(z\sim0.1)$. The structured parameters of the jet are the same as in \ref{fig:spec_con&inf} and \ref{fig:spec_con&fin}. The luminosity indices are $a_{\mathrm r}=0.75$, $a_{\mathrm d}=-2$, $t_{\mathrm p}=2.4$~s. Different colors stand for different $t_{\mathrm{obs}}$, 1~s (magenta), 3~s (red), 5~s (cyan), 10~s (green), and 15~s (deepskyblue). Different shaded areas indicate the detectable flux range of SVOM-ECLAIRs (4-250~keV, orange), Swift-BAT (15-150~keV, limegreen), EP-WXT (0.5-4~keV, dodgerblue)}
    \label{fig:spec_var&fin}
\end{figure*}

\begin{figure*}
    \centering
    \includegraphics[width=\textwidth]{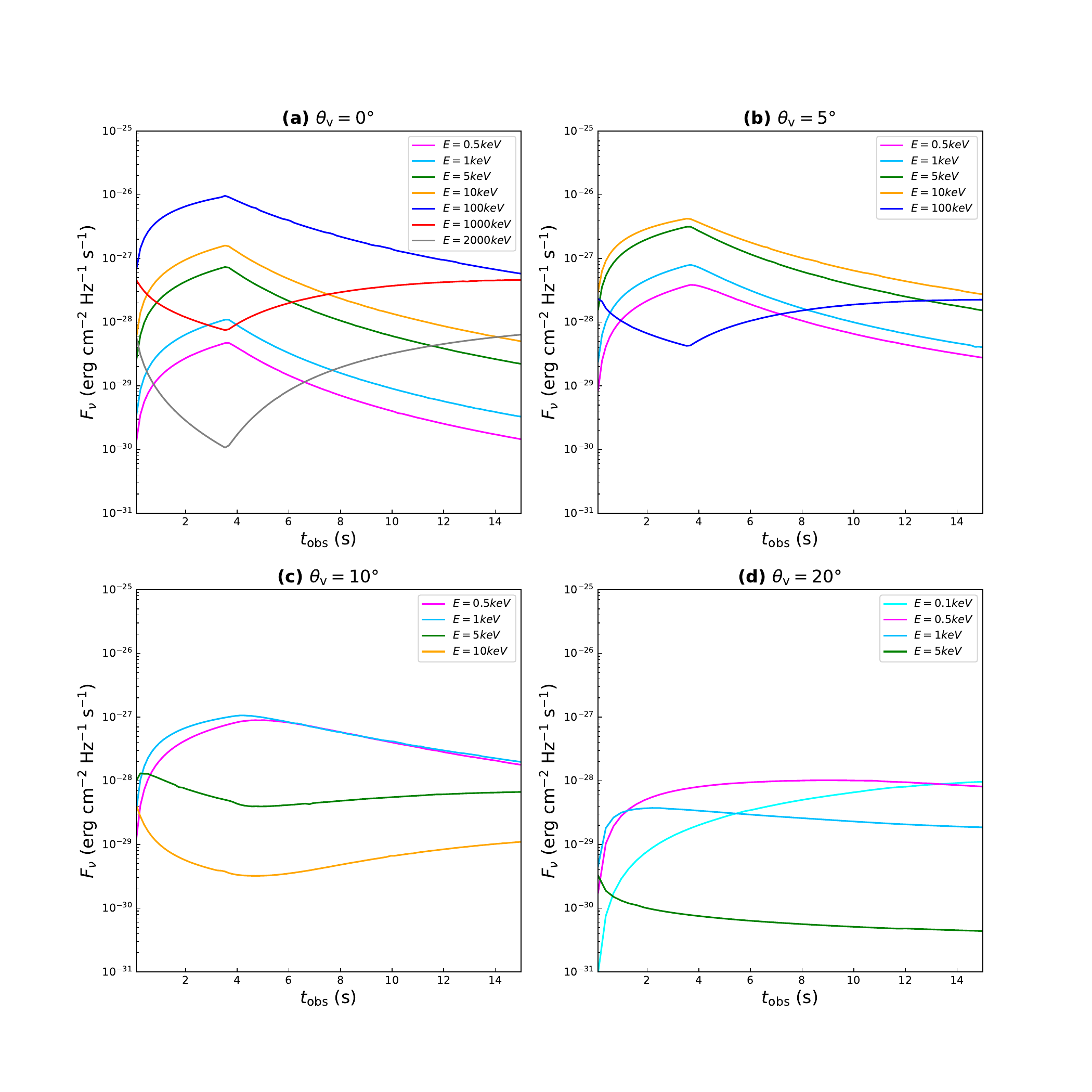}
    \caption{Light curve with variable luminosity and finite outer boundary by different viewing angles $\theta_{\mathrm v}$. Parameters are the same as the solid spectrum in Figure \ref{fig:spec_var&fin}, different colors stand for different energies, 0.5~keV (magenta), 1~keV (deepskyblue), 5~keV (green), 10~keV (orange), 100~keV (b), 1000~keV (red), 2000~keV (gray).}
    \label{fig:cur_var&fin}
\end{figure*}

The light curve of the prompt pulses could be approximated by the broken power law \citep{Salvaterra_2012, Pescalli_2016}. Assuming that the corresponding peak luminosity of the inner core is $L_{\mathrm{cp}}$, the peak time is $t_{\mathrm{p}}$, and the rising and decaying indices are $a_{\mathrm{r}}$ and $a_{\mathrm{d}}$, the inner core luminosity history could thus be expressed as
\begin{equation}
    L_{\mathrm{c}}(\hat{t})=\left\{
    \begin{aligned}
        &10^{a_{\mathrm{r}} \log \hat{t}\,+\,b_{\mathrm{r}}}~\mathrm{erg}\: \mathrm{s}^{-1}, && \text{if}\ \ 0 < \hat{t} \leq t_{\mathrm{p}}, \\
        &10^{a_{\mathrm{d}} \log \hat{t}\,+\,b_{\mathrm{d}}}~\mathrm{erg}\: \mathrm{s}^{-1}, && \text{if}\ \ \hat{t} > t_{\mathrm{p}},
    \end{aligned}
    \right.
\end{equation}
where $b_{\mathrm{r}}=\log L_{\mathrm{cp}}-a_{\mathrm{r}} \log \hat{t}_{\mathrm{p}}$
and $b_{\mathrm{d}}=\log L_{\mathrm{cp}}-a_{\mathrm{d}} \log \hat{t}_{\mathrm{p}}$.
The instantaneous photosphere spectra under the assumption of the finite boundary with variable luminosity can also be obtained by integrating Equation \ref{eq:lay-contr} over central engine time $\hat{t}$. The instantaneous spectra for different viewing angles are shown in Figure \ref{fig:spec_var&fin}, and the light curves are shown in Figure \ref{fig:cur_var&fin}. It is obvious that the peak flux of the instantaneous spectra rises as the luminosity increases and then falls as the luminosity decreases. 

The low energy light curves show a rise to a maximum followed by a decline, which is similar to the trend in luminosity, while at higher energy the light curves first decline and then rise (Figure \ref{fig:cur_var&fin}).
This indicates that the spectrum evolves from hard to soft as the luminosity increases, but evolves from soft to hard as the luminosity decreases, which is consistent with the conclusions obtained in \cite{Deng_2014}. 
As we mentioned in Section \ref{subsec:con&inf}, the evolution of the spectrum slows down significantly for larger viewing angles, so the spectrum is still evolving from hard to soft at $t_{\mathrm{obs}}=15$~s, which is significantly delayed than $(1+z)t_{\rm p}$.

EP-WXT, Swift-BAT, and SVOM-ECLAIRs will all have the capability to detect the quasi-thermal photosphere emission of GRBs in both on-axis and off-axis scenarios. In addition, according to our results, bright bursts, if dominated by quasi-thermal emission, will exhibit lower peak energies, which are closer to the detection energy band like EP-WXT (0.5–4~keV), thereby making EP-WXT effective in detecting photosphere emission from bright bursts. By incorporating the GRB luminosity function with the local rate derived from recent studies~\citep{Hou_2018,Lan_2019,Lan_2021,Gao_2025}, we may estimate the detection rates for current detectors.
Assuming that a fraction of GRBs may have thermal emissions, SVOM-ECLAIRs enhances our detection capability for GRBs' photosphere emission (roughly twice that of Swift-BAT).
For EP-WXT, its wide field of view enables a higher detection rate of long GRBs, of which a majority could be viewed off-axis~\citep{Gao_2024}.
Thus, EP would also increases the possibility to capture the GRB thermal emission.


\subsection{Detectability Of Photosphere Emission From Short GRBs}\label{subsec:det}
\begin{figure}
    \centering
    \includegraphics[width=0.5\textwidth]{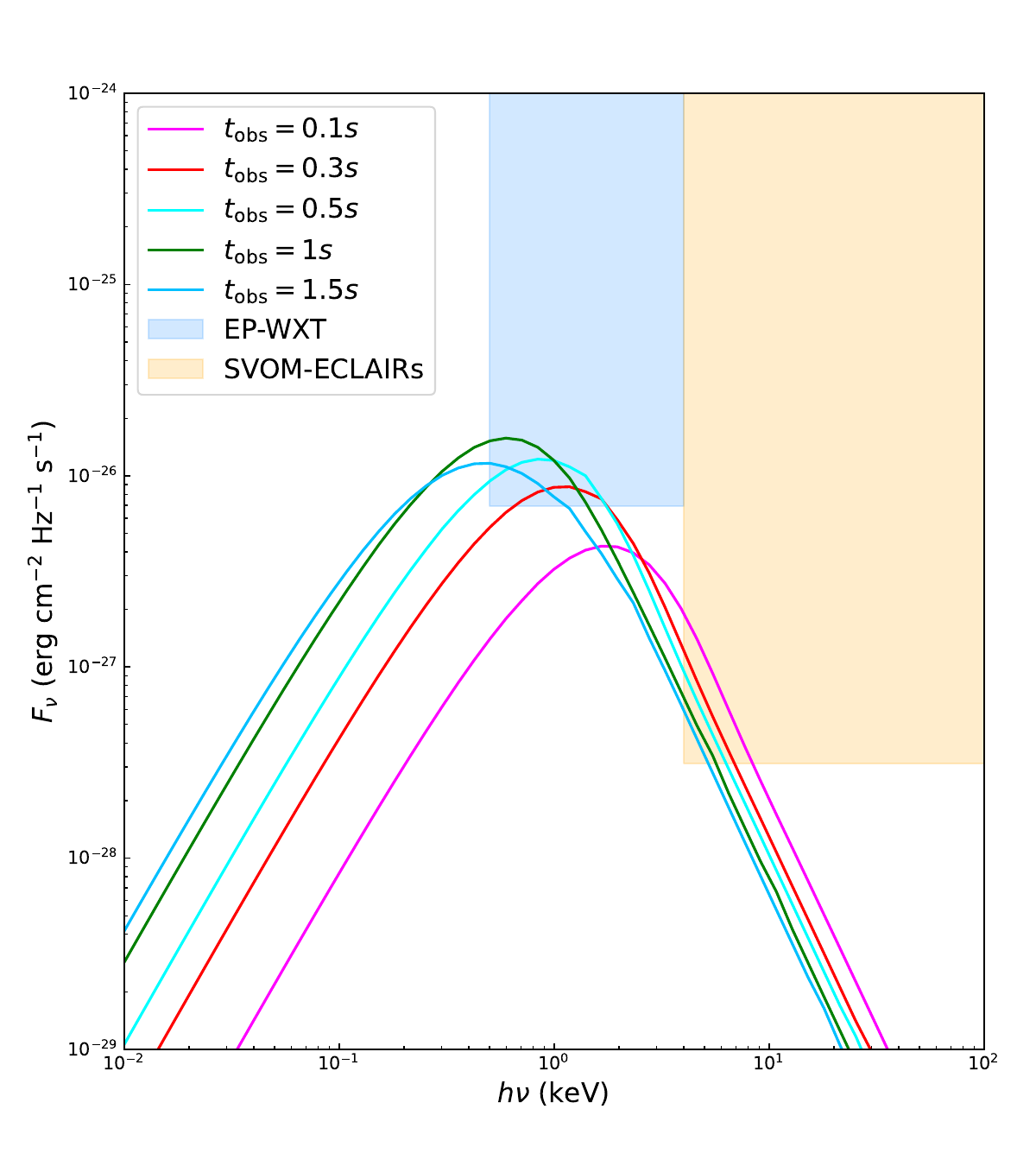}
    \caption{Theoretical instantaneous photosphere spectra of short GRB similar to GRB 170817A by $\theta_{\mathrm{v}}=27.6\degree$. Parameters are $\Gamma_{\mathrm c}=100$, $L_{\mathrm c}=10^{51}~\mathrm{erg\:s^{-1}}$, $d_{\mathrm L}=1.236\times10^{26}~\mathrm{cm}$ ($\sim 40~\mathrm{Mpc}, z=0.009$), $\theta_{\mathrm c}=5.1\degree$, $\theta_{\mathrm m}=30\degree$, index values are $k_{\mathrm{\Gamma}}=2$, $k_{\mathrm e}=4.3$, and the luminosity indices are $a_{\mathrm r}=0.75$, $a_{\mathrm d}=-2$, $t_{\mathrm p}=0.3$~s, duration of the burst is $t_{\mathrm d}=1.5$~s. Different colors stand for different $t_{\mathrm{obs}}$, $0.1$~s (magenta), $0.3$~s (red), $0.5$~s (cyan), $1$~s (green), $1.5$~s (deepskyblue). The blue and orange shaded area indicates the detectable flux range of EP-WXT (0.5-4~keV) and SVOM-ECLAIRs (4-250~keV).}
    \label{fig:spec_sGRB}
\end{figure}

There is some evidence for photosphere emissions in the prompt phase of GRB 170817A~\citep{Meng_2018, Gottlieb_2018}. We calculate the expected photosphere emission of short GRB similar to GRB 170817A using the variable luminosity with a finite outer boundary model. Taking the jet structure and the viewing angle
of GRB 170817A \citep{Li_2019}, a group of instantaneous photosphere spectra is obtained (see Figure \ref{fig:spec_sGRB}). This indicates that EP-WXT and SVOM-ECLAIRs possess the capability to detect off-axis high-energy
transient sources similar to GRB 170817A.

\begin{figure*}
    \centering
    \includegraphics[width=\textwidth]{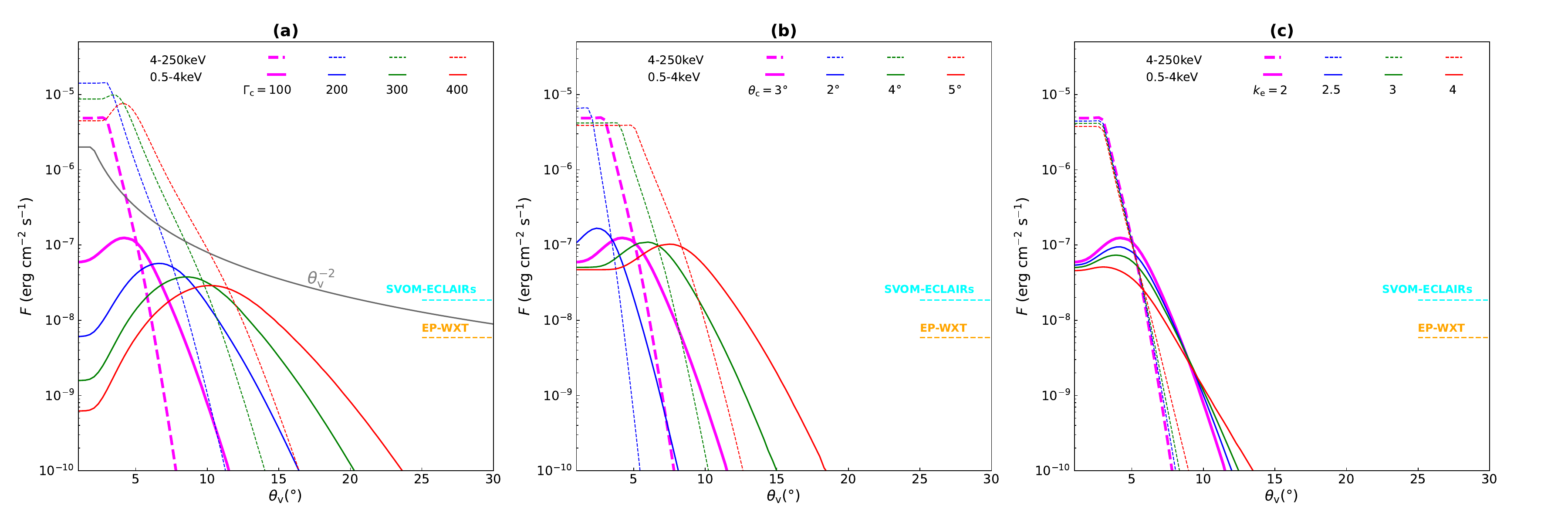}
    \caption{This figure shows the dependence of the detection flux on the viewing angle for different structured parameters, the magenta-colored bold curves in the three subfigures have the same parameters of $\Gamma_{\mathrm c}=100$, $L_{\mathrm c}=10^{51}\mathrm{erg\:s^{-1}}$, $d_{\mathrm L}=6.17\times10^{26}\mathrm{cm}$ ($\sim 200\mathrm{Mpc}, z=0.045$), $\theta_{\mathrm c}=3\degree$, $\theta_{\mathrm m}=30\degree$, $k_{\mathrm{\Gamma}}=2$, $k_{\mathrm e}=2$. Subfigures (a) (b) (c) show the detected flux at different viewing angles under different $\Gamma_{\mathrm c}$, $\theta_{\mathrm c}$, $k_{\mathrm e}$, respectively. The dashed line on the right indicates the detectable flux range of EP-WXT (0.5-4keV, orange), and SVOM-ECLAIRs (4-250keV, cyan), grey line indicates the angular dependence of isotropic luminosity.}
    \label{fig:flux_EP&SVOM}
\end{figure*}

In general, the short GRB viewed at a relatively large off-axis angle is faint.
However, the exact jet core angle and the jet structure, both poorly constrained for most bursts, could affect its detection rate. We then further explored the structured parameters to analyze the detectability of typical short GRBs viewed off-axis. 
As a fiducial calculation, we assume a short burst at a distance of 200~Mpc ($z\sim0.045$), with a duration of 1.5 s and a constant luminosity equal to the typical value of $10^{51}~\mathrm{erg\:s^{-1}}$ for short burst luminosity. The detected flux can be calculated by:
\begin{equation}
    F(\theta_{\mathrm{v}})=\frac{1+z}{{d_{\mathrm L}^{2}}}
\iiint\frac{d\dot{N}(\theta_{\mathrm j})}{d \Omega}P(r,\Omega)R(\nu,T)h\nu d\Omega dr d\nu_{\mathrm{obs}}.  \label{eq:EP_det} 
\end{equation}
The dependence of the detection flux on the viewing angle for two detectors, EP-WXT and SVOM-ECLAIRs, are shown in Figure \ref{fig:flux_EP&SVOM}.

It is found that the photosphere emission of such a typical short burst is very likely to be detected by EP-WXT and SVOM-ECLAIRs within the viewing angle of $10\degree$.
Comparing the luminosity-angle dependence with the detected flux (Figure \ref{fig:flux_EP&SVOM}(a)), we found that the dependence of detection flux on the viewing angle is steeper than the luminosity dependence on the viewing angle ($\propto \theta_{\mathrm{v}}^{-2}$).
This is due to the fact that there is not only a lower luminosity in the off-axis direction but also a lower Lorentz factor, which leads to a greater optical depth (Equation \ref{eq:tau_con&fin}), so the detection flux decreases more rapidly with the increase of viewing angle.

The Lorentz factor of the inner-core $\Gamma_{\mathrm{c}}$ also affects the flux-angle dependence. As a larger Lorentz factor corresponds to a smaller optical depth, the detection flux decreases more shallowly with the increase of the viewing angle. 

Figure \ref{fig:flux_EP&SVOM} also shows that the ``optimal detection angle'' for EP-WXT is structure dependent rather than on-axis. This is an joint effect effect of the jet structure and the limited band of the instrument.
The temperature of emissions near the photosphere radius is $T(r_{\mathrm{ph}},\theta_{\mathrm{v}})=T_{0}(\theta_{\mathrm v})(r_{\mathrm{ph}}/{r_{\mathrm s}})^{-2/3}$.
Combining Equations \ref{eq:tem}, \ref{eq:cen-tem}, \ref{eq:rs}, \ref{eq:rph_appro},
this temperature corresponds to the spectral peak energy of
\begin{align}
    &E_{\mathrm{peak}}\simeq\notag\\
    &\left\{
    \begin{array}{l}
    \frac{14}{1+z}~(\frac{L_{\mathrm{c}}}{10^{52}})^{-\frac{5}{12}}(\frac{\Gamma_{\mathrm{c}}}{100})^{\frac{8}{3}}(\frac{r_{0}}{10^7})^{\frac{1}{6}}~\mathrm{keV}, \text{\ }\theta_{\mathrm{v}}\leq\theta_{\mathrm{c}},\\
    \frac{14}{1+z}~(\frac{L_{\mathrm{c}}}{10^{52}})^{-\frac{5}{12}}(\frac{\Gamma_{\mathrm{c}}}{100})^{\frac{8}{3}}(\frac{r_{0}}{10^7})^{\frac{1}{6}}(\frac{\theta_{\mathrm{v}}}{\theta_{\mathrm{c}}})^{\frac{5}{12}k_{\mathrm{e}}-\frac{8}{3}k_{\Gamma}}~\mathrm{keV},\text{\ }\theta_{\mathrm{c}}<\theta_{\mathrm{v}}\leq\theta_{\mathrm{m}}.  
    \end{array}
    \right.   \label{eq:Ep_appro} 
\end{align}
The EP-WXT only gathers the energy on a narrow band (0.5-4~keV), which is only a small fraction of the total energy for thermal radiation. 
According to Equation \ref{eq:Ep_appro}, although $E_{\rm peak}$ is lower for larger $\theta_{\rm v}$, it is closer to the WXT's band, resulting in a larger flux level. Also, the optimal detection angle increases with $\Gamma_{\rm c}$ and $\theta_{\rm c}$ from Equation \ref{eq:Ep_appro}. 
However, it is insensitive to values of $k_{\rm e}$ in Figure \ref{fig:flux_EP&SVOM}(c). This is because that the detected flux at different viewing angles is strongly modulated by the optical depth in that direction.

\section{CONCLUSIONS AND DISCUSSION} \label{sec:conclu}

In this paper, we explored the characteristics of photosphere emission from the prevailing structured jet with detailed numerical calculations.
In general, the early instantaneous spectrum at the low-frequency range below the peak energy is close to the blackbody spectrum under the Rayleigh-Jeans approximation ($F_{\nu}\propto \nu^{2}$), and the spectral index will then decreases slightly and approach 1.5 gradually ($F_{\nu}\propto \nu^{1.5}$), caused by the superposition of blackbody spectra with multi-temperatures.
Both the spectral peak flux and the peak energy decreases with the viewing angle $\theta_{\rm v}$. Meanwhile, a larger viewing angle would result in a lower evolving rate of the instantaneous spectrum, which is caused by the angular dependence of the photosphere radius. In comparison with the on-axis case, the larger photosphere radius at off-axis leads a longer saturation timescale due to the longer time delay of radiation to the observer.

The early emission is sensitive to the treatment of the outflow boundary.
We demonstrate that the infinite approximation of the boundary will underestimate the observed flux at the early stage. 
For the realistic case of finite boundary, the early emission with higher temperature is easy to escape, resulting in a ``hard'' component in the high frequencies.
In the off-axis case, this component sustains for a longer time when the viewing angle is large enough, making it distinguishable from observations.

When the luminosity of the central engine is variable, $F_{\nu, \mathrm{max}}$ track with luminosity over time, while $E_{\rm peak}$ tracks conversely with luminosity over time. The spectrum evolves from ``hard-to-soft'' in the rising phase but evolves from ``soft-to-hard'' in the decaying phase, which is consistent with those of uniform jets in \cite{Deng_2014}. However, the relatively slower spectral evolving rate for an off-axis observer would smooth the light curves.  

Detecting photosphere emissions in the current era is promising.
For a short burst with a similar structure to GRB 170817A and a typical luminosity of $10^{51}~\mathrm{erg\:s^{-1}}$ within a distance of 200~Mpc, it is possible for the EP-WXT and SVOM-ECLAIRs to detect it within a viewing angle of $10\degree$.
In addition, the flux-angle dependence does not follow the structure of the jet luminosity directly but is significantly related to the Lorentz factor and the inner-core half-opening angle. For cases of larger Lorentz factor and core angle, the dependence of the observed flux on $\theta_{\rm v}$ is more shallow.

The assumption of saturation regime ($r_{\mathrm s} < r_{\mathrm{ph}}$) is adopted in our work. However, since the jet luminosity along specific angles has a broad distribution range, an unsaturated situation may hold for some parts of the jet. Therefore, it is necessary to consider the bulk acceleration in the unsaturated situation in further work. Also, the energy exchange between the photons and the plasma during the scatterings should be incorporated.

While the traditional quasi-thermal photosphere emission model may have issues in explaining the low-energy spectral slope, the Comptonized photosphere model proposes that the observed Band-like spectrum could originate from photospheric radiation transformed through sub-photospheric processes~\citep{Meszaros_2000,Rees_2005,Pe'er_2006,Beloborodov_2010,Ryde_2010,Vurm_2011}. In addition to the quasi-thermal component, GRBs may contain multiple radiation components such as synchrotron emission from shocks or magnetic reconnection~\citep{Burgess_2020,Geng_2018_apjs,Gao_2021}. Our calculations indicate that the quasi-thermal component becomes more detectable during early radiation phases, particularly in off-axis observation scenarios. When multiple radiation components coexist, observational data alone cannot conclusively determine the dominance of quasi-thermal radiation, necessitating comprehensive model fitting for component discrimination. Therefore, a more generic physical model needs to be developed, self-consistently incorporating the Comptonized photosphere structure, synchrotron radiation mechanisms, jet propagation dynamics, and energy dissipation processes via internal shocks~\citep{Samuelsson_2023, Alamaa_2024} or magnetic reconnection, with consideration of synergistic interactions among multiple radiation components. 

\begin{acknowledgments}
We appreciate valuable comments and suggestions from the anonymous referee. We thank Di Xiao, Jun-Jie Wei, Ye Li for helpful discussions or comments.  This study is partially supported by the National Natural Science Foundation of China (grant Nos. 12273113, 12321003, 12393812, and 12393813), the International Partnership Program of Chinese Academy of Sciences for Grand Challenges (114332KYSB20210018). J.J.G. acknowledges support from the Youth Innovation Promotion Association (2023331). H.X.G. acknowledges support from Jiangsu Funding Program for Excellent Postdoctoral Talent.
\end{acknowledgments}

\vspace{5mm}
\newpage
\appendix 

\section{The Probability Density Function Of The Last Scattering Event}\label{app:prob}
Considering the radiation transferring in a medium with an optical depth of $\tau$, a beam of light with an initial intensity of $I_{0}$ passing through this medium will experience an exponential attenuation, resulting in an outgoing intensity of $I_{0}e^{-\tau}$. From a probability perspective, we can assume that the probability of the photon passing through the medium without being scattered is $e^{-\tau}$, and the probability of being scattered is $1-e^{-\tau}$ ($1-e^{-\tau}\sim\tau$ when $\tau \ll 1$). Therefore, the probability of a photon being scattered within the segment $r-r+dr$ is \citep{Deng_2014}
\begin{equation}
    P_{r} dr=d\tau=D^{-1}\sigma_{\mathrm T}n^{\prime}dr=\frac{\sigma_{\mathrm T}n}{2\Gamma^{2}}dr.
\end{equation}
In the comoving frame, the probability of a photon being scattered in any direction is written as
\begin{equation}
    P^{\prime}_{\mathrm{\Omega^{\prime}}}d\Omega^{\prime}\propto \frac{d\Omega^{\prime}}{4\pi},
\end{equation}
while in the observer frame, the probability density of a photon being scattered towards a specific solid angle $\Omega$ is \citep{Pe'er_2008}:
\begin{equation}
    P_{\mathrm{\Omega}}d\Omega\propto\frac{D^{2}d\Omega}{4\pi}.
\end{equation}
Also, the probability of this scattered photon has not been scattered before is $e^{-\tau}$. All these conditions result in the probability density distribution of the last scattering event to be
\begin{equation}
    P(r,\Omega)drd\Omega \propto P_{r} dr\cdot P_{\mathrm{\Omega}}d\Omega\cdot e^{-\tau},
\end{equation}
namely
\begin{equation}
    P(r,\Omega)=P(r,\theta,\phi)=\frac{D^{2}(\theta,\phi)\sigma_{\mathrm T}n(r,\theta,\phi)e^{-\tau}}{8\pi\Gamma^{2}(\theta,\phi) A},\label{eq:app_prob}
\end{equation}
where A is the proportionality constant.

\section{Luminosity Distribution Function}\label{app:lum}
In the assumption of variable luminosity, the distribution of luminosity varies with both angles and time, it can be expressed as
\begin{equation}
    L(\theta_{\mathrm{j}},\hat{t})=\left\{
    \begin{aligned}
        &10^{a_{\mathrm{r}} \log\hat{t}\,+\,b_{\mathrm{r}}} \mathrm{erg~s}^{-1}, && \text{if}\ \ 0 < \hat{t} \leq t_{\mathrm{p}},\theta \leq \theta_{\mathrm{c}}, \\
        &10^{a_{\mathrm{r}} \log\hat{t}\,+\,b_{\mathrm{r}}}(\theta /\theta_{\mathrm c})^{-k_{\mathrm e}} \mathrm{erg~s}^{-1}, && \text{if}\ \ 0 < \hat{t} \leq t_{\mathrm{p}},\theta_{\mathrm{c}}<\theta \leq \theta_{\mathrm{m}}, \\
        &10^{a_{\mathrm{d}} \log\hat{t}\,+\,b_{\mathrm{d}}} \mathrm{erg~s}^{-1}, && \text{if}\ \ \hat{t} > t_{\mathrm{p}},\theta \leq \theta_{\mathrm{c}},\\
        &10^{a_{\mathrm{d}} \log\hat{t}\,+\,b_{\mathrm{d}}}(\theta /\theta_{\mathrm c})^{-k_{\mathrm e}} \mathrm{erg~s}^{-1}, && \text{if}\ \ \hat{t} > t_{\mathrm{p}},\theta_{\mathrm{c}}<\theta \leq \theta_{\mathrm{m}}, \\
        & 0 \ \mathrm{erg~s}^{-1}, &&\text{else.}
    \end{aligned}
    \right.
\end{equation}

\section{Calculation Of Optical Depth Under Variable Luminosity}
\begin{figure}
    \centering
    \includegraphics[width=0.7\textwidth]{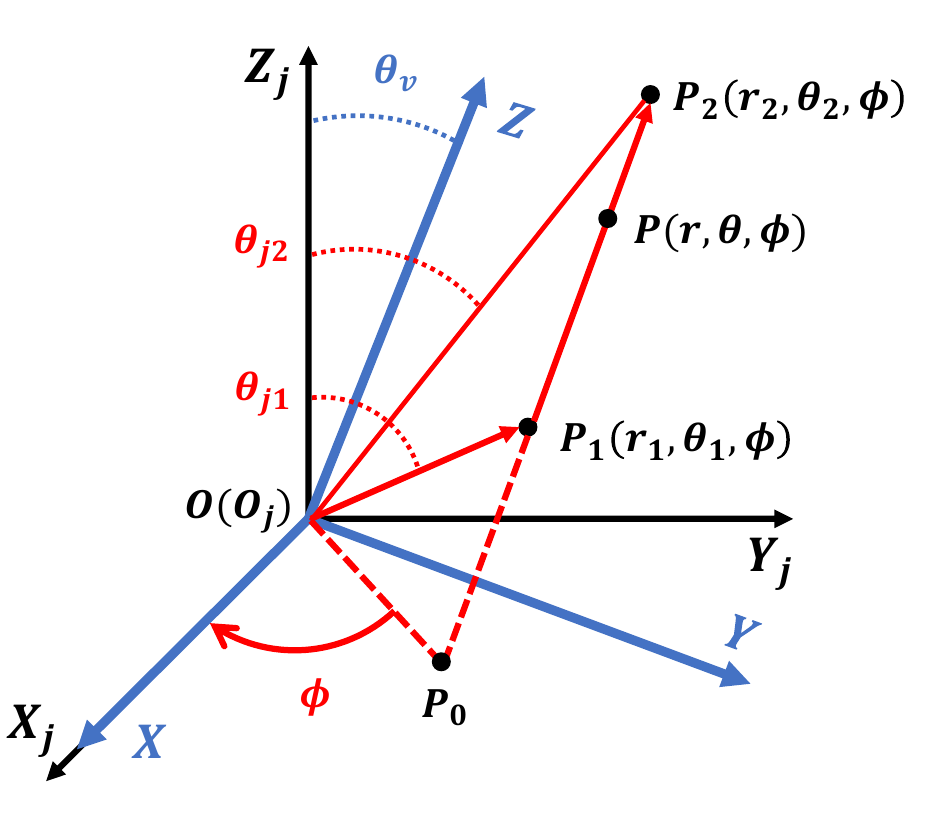}
    \caption{The geometric diagram of the optical depth calculation. The diagram shows two coordinate systems with overlapping origins and the X-axis, where the origin represents the central engine. The coordinate system $X_{\mathrm j}Y_{\mathrm j}Z_{\mathrm j}$ represents the jet frame, with the $Z_{\mathrm j}$-axis being the jet axis (axis of symmetry), while the XYZ represents the observer frame, with the Z-axis pointing towards the observer. The angle between the Z-axis and $Z_{\mathrm j}$-axis is the viewing angle $\theta_{\mathrm{v}}$. The photon scatters for the last time and escapes at point $P_{1}(r_{1}, \theta_{1},\phi)$, and then catches up with the outer boundary at point $P_{2}(r_{2}, \theta_{2},\phi)$. $P_{0}$ is the projection of line segment $P_{1}P_{2}$ onto the XOY plane, $P(r,\theta,\phi)$ is an arbitrary point on the line segment $P_{1}P_{2}$, $\theta_{\mathrm{j1}}$ and $\theta_{\mathrm{j2}}$ represent the polar angles of $P_{1}$ and $P_{2}$ in the jet frame. }
    \label{fig:cal_tau}
\end{figure}
\label{app:cal_tau}
As shown in Figure \ref{fig:cal_tau}, the probability of a photon injected into the photosphere at $\hat{t}$ undergoing its last scattering at $P_{1}$ depends on the optical depth along the segment $P_{1}P_{2}$. Since luminosity varies with time, the spatial distribution of plasma density also varies over time. If we can determine the comoving plasma density $n$ at any point $P$ along the line segment $P_{1}P_{2}$ when the photon passes through during the escape process, we can calculate the optical depth of this segment using Equation \ref{eq:tau_con&fin}. The length of line segment $PP_{1}$ is
\begin{equation}
    \lvert PP_{1} \rvert = r\cos\theta-r_{1} \cos\theta_{1}.
\end{equation}
We can consider this process as the encounter between electrons moving along the radial direction and photons escaping along the line of sight at point $P$. Before this encounter, the radius at which these electrons are located when the photon undergoes its last scattering at $P_{1}$ is
\begin{equation}
    r^{\prime}=r-\beta(\theta_{\mathrm{j}})c\times\frac{\lvert PP_{1} \rvert}{c}=r-\beta(\theta_{\mathrm{j}})(r\cos\theta-r_{1}\cos\theta_{1}).
\end{equation} 
Therefore, the comoving plasma density at point $P$ when the photon arrives can be expressed as
\begin{equation}
    n^{\prime}(r,\theta)=\frac{\dot{M}(\theta_{\mathrm j},\hat{t}-[\frac{r^{\prime}}{\beta(\theta_{\mathrm j})c}-\frac{r_{1}}{\beta(\theta_{\mathrm{j1}})c}])}{4\pi m_{\mathrm{p}}\beta(\theta_{\mathrm j})c\Gamma(\theta_{\mathrm j})r^2}=\frac{L(\theta_{\mathrm j},\hat{t}-[\frac{r^{\prime}}{\beta(\theta_{\mathrm j})c}-\frac{r_{1}}{\beta(\theta_{\mathrm{j1}})c}])}{4\pi m_{\mathrm{p}}\beta(\theta_{\mathrm j})c^3\Gamma^2(\theta_{\mathrm j})r^2}.
\end{equation}
The $\theta_{\mathrm{j}}$ represents the polar angle of $P$ in the jet frame, and the relationship between $\theta$ and $\theta_{\mathrm j}$ is given by Equation \ref{eq:co_trans}. Therefore, the optical depth under the assumption of variable luminosity can be calculated using 
\begin{equation}
   \tau=\int_{r_{1}}^{r_{2}}D^{-1}\sigma_{\mathrm T}n^{\prime}\frac{dr}{\cos\theta}=\int_{r_{1}}^{r_{2}}\frac{L(\theta_{\mathrm j},\hat{t}-[\frac{r^{\prime}}{\beta(\theta_{\mathrm j})c}-\frac{r_{1}}{\beta(\theta_{\mathrm{j1}})c}])}{4\pi m_{\mathrm{p}}\beta(\theta_{\mathrm j})c^3\Gamma(\theta_{\mathrm j})r^2}\cdot\sigma_{\mathrm T}(1-\beta(\theta_{\mathrm j})\cos\theta)\cdot\frac{dr}{\cos\theta}. 
\end{equation}

\section{Optical Depth And Photosphere Radius}\label{app:rph}
The expression for optical depth is
\begin{equation}
        \tau=\int_{r_{1}}^{r_{2}}D^{-1}\sigma_{\mathrm T}n^{\prime}\frac{dr}{\cos(\theta)}=\int_{r_{1}}^{r_{2}}\frac{\dot{M}(\theta_{\mathrm j})\sigma_{\mathrm{T}}}{4\pi m_{\mathrm{p}}\beta(\theta_{\mathrm j})c\Gamma(\theta_{\mathrm j})r^2}\Gamma(\theta_{\mathrm{j}})[1-\beta(\theta_{\mathrm j})\cos(\theta)]\frac{dr}{\cos(\theta)}, 
\end{equation}
during the photon escape, the conserved quantity is
\begin{equation}
    r\sin(\theta)=d=const,
\end{equation}
converting the integrals according to the above conservation relation, we can further obtain
\begin{equation}
    \tau=\int_{\theta_{2}}^{\theta_{1}}\frac{d\dot{M}(\theta_{\mathrm j})}{d\Omega}\frac{\sigma_{\mathrm{T}}}{m_{\mathrm{p}}cr^2}\frac{1-\beta(\theta_{\mathrm j})\cos(\theta)}{\beta(\theta_{\mathrm{j}})}\frac{r}{\sin(\theta)}d\theta=\frac{\sigma_{\mathrm{T}}}{m_{\mathrm{p}}c}\frac{1}{r_{1}\sin(\theta_{1})}\int_{\theta_{2}}^{\theta_{1}}\frac{d\dot{M}(\theta_{\mathrm j})}{d\Omega}\frac{1-\beta(\theta_{\mathrm j})\cos(\theta)}{\beta(\theta_{\mathrm{j}})}d\theta,
\end{equation}
combined with the definition of the photosphere radius in Equation \ref{eq:rph}, we can obtain an integral expression for the photosphere radius
\begin{equation}
    r_{\mathrm{ph}}=\frac{\sigma_{\mathrm{T}}}{m_{\mathrm{p}}c}\frac{1}{\sin(\theta_{1})}\int_{\theta_{2}}^{\theta_{1}}\frac{d\dot{M}(\theta_{\mathrm j})}{d\Omega}\frac{1-\beta(\theta_{\mathrm j})\cos(\theta)}{\beta(\theta_{\mathrm{j}})}d\theta,
\end{equation}
so the relationship between optical depth and photosphere radius is
\begin{equation}
    r_{\mathrm{ph}}=\tau \times r_{1}.
\end{equation}

\bibliography{main}{}
\bibliographystyle{aasjournal}

\end{document}